\documentclass[twocolumn,footinbib,notitlepage,showpacs,amsmath,amssymb,pra,aps,10pt]{revtex4-1}
\usepackage{units}
\usepackage[pdftex]{graphicx}              
\usepackage[pdftex]{hyperref}
\usepackage{pdfsync}
\usepackage{bm}	

\hypersetup{pdfauthor=Sherlock and Gildemeister, pdftitle= Evaporative Cooling by Landau-Zener losses, colorlinks=true, linkcolor=black, citecolor=black, filecolor=black, urlcolor=black, bookmarksopen = False, bookmarksnumbered = true, linktocpage = true, pdfstartview = FitV, pdfstartpage=1}
 
\begin{document}
\title{Novel techniques to cool and rotate Bose-Einstein condensates in time-averaged adiabatic potentials}

\author{M. Gildemeister}
\author{B. E. Sherlock}
\author{C. J. Foot}
\affiliation{Clarendon Laboratory, University of Oxford, Parks Road, Oxford, OX1 3PU, United Kingdom}

\date{\today}

\begin{abstract}
We report two novel techniques for cooling and rotating Bose-Einstein condensates in a dilute rubidium vapour that highlight the control and versatility afforded over cold atom systems by time-averaged adiabatic potentials (TAAPs). The intrinsic loss channel of the TAAP has been successfully employed to evaporatively cool a sample of trapped atoms to quantum degeneracy. The speed and efficiency of this process compares well with that of conventional forced rf-evaporation. In an independent experiment, we imparted angular momentum to a cloud of atoms forming a Bose-Einstein condensate by introducing a rotating elliptical deformation to the TAAP geometry. Triangular lattices of up to 60 vortices were created. All findings reported herein result from straightforward adjustments of the magnetic fields that give rise to the TAAP.
\end{abstract}

\maketitle

\section{Introduction}
\label{introd}
The precision with which atomic motion can be controlled and manipulated in conservative trapping potentials was a strong motivation for the early development of cold atom research. Today the myriad of physical phenomena that are studied with cold atoms makes the requirement for precise control  more relevant than ever. In this article, we report on recent experimental results of ultracold atoms trapped in time-averaged adiabatic potentials (TAAPs) that demonstrate the unique combination of control and versatility afforded by this trap.

The TAAP was first proposed in 2007 \cite{vonklitzing2007} and has since been experimentally shown to possess many attractive features for working with cold atoms. The TAAP emulates the simplicity and robustness of operation of magnetostatic traps whilst providing access to a broad range of geometries and dynamic manipulations. The positioning of the magnet coil several centimeters from the atoms gives rise to very smooth potentials that allow the preservation of the atomic phase coherence during changes to the confinement. In the first successful trapping of atoms in a TAAP trap reported in \cite{marcus}, cold atoms were loaded into a vertically offset double-well potential which could be adiabatically transformed into an anisotropic shell potential. Building on this work a successful technique was developed to trap atoms in a horizontally orientated, ring-shaped TAAP with a continuously adjustable ring radius \cite{ben}. 

In this article, experimental results demonstrating novel techniques to separately cool and rotate a cold atom cloud in a TAAP are presented. Efficient evaporative cooling across the Bose-Einstein condensate (BEC) phase transition is realised via the Landau-Zener (LZ) loss channel. Evidence for cooling via this route has previously been reported \cite{perrin2007}, however the presence of non-negligible heating in the trap and relatively low elastic collision rates resulted in only modest increases in phase space density. By contrast the experimental results presented here show a strong and rapid cooling across the phase transition with an efficiency that compares well with conventional rf forced evaporation. We demonstrate the nucleation of a vortex lattice in a BEC trapped in a TAAP by dynamically adjusting the geometry such that a rotating elliptical deformation can be introduced to the trap. 

This article is organised as follows: Sections \ref{taaptheory} and \ref{exp} review the theoretical basis for the TAAP and describe the experimental apparatus with which all of the data presented herein were taken (much of the content has been reported in detail in previous publications \cite{marcus, ben}). In Sec.~\ref{natevap} the results of natural evaporation in the double-well TAAP trap are presented. Section \ref{vortexnuc} includes an explanation of the novel TAAP rotation scheme and the relevant experimental results. The conclusions in Sec. \ref{conc} give an outlook on future work that has been stimulated by these results. 
\section{\label{taaptheory}TAAP Theory}

\begin{figure*}[t]
\centering
\includegraphics[width=1.8\columnwidth]{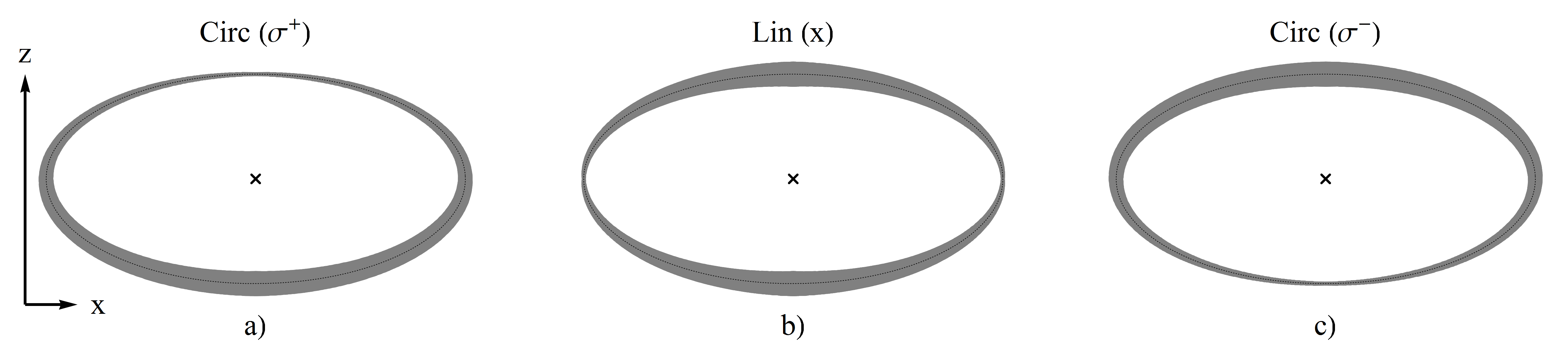}
\caption{Variation of $\Omega_{\mathrm{R}}$ around a cross-section of the resonant ellipsoid (dotted black line) in the $y=0$ plane. The thickness of the line (grey shading) represents the magnitude of $\Omega_{\mathrm{R}}$ i.e. it is thickest where $\Omega_{\mathrm{R}}$ has a maximum value. Three polarisations are shown: a) circular $\bm{\mathrm{B}}_{\mathrm{rf}}(\bm{r},t) = B_{\mathrm{rf}}[\cos{(\omega_{\mathrm{rf}}t)\hat{\bm{e}}_{x}}+\sin{(\omega_{\mathrm{rf}}t)}\hat{\bm{e}}_{y}]$, b) linear $\bm{\mathrm{B}}_{\mathrm{rf}}(\bm{r},t) = B_{\mathrm{rf}}\cos{(\omega_{\mathrm{rf}}t)\hat{\bm{e}}_{x}}$, and c) circular with opposite handedness to that in a) $\bm{\mathrm{B}}_{\mathrm{rf}}(\bm{r},t) = B_{\mathrm{rf}}[\cos{(\omega_{\mathrm{rf}}t)\hat{\bm{e}}_{x}}-\sin{(\omega_{\mathrm{rf}}t)}\hat{\bm{e}}_{y}]$.}
\label{polarisations}
\end{figure*}

A TAAP is formed by the superposition of three independent magnetic fields which can be categorised by their time dependence. The underlying framework is provided by a spatially inhomogeneous, static magnetic field $\bm{\mathrm{B}}_{\mathrm{q}} (\bm{\mathrm{r}})$. The addition of radio-frequency field $\bm{\mathrm{B}}_{\mathrm{rf}}(\bm{r},t)$ (spatially uniform, oscillating in the MHz realm) transforms the magnetostatic potential into an adiabatic potential (AP) by strongly driving transitions between Zeeman sub-levels of the atom \cite{zobay2001}. The time-averaging of the AP is performed by a magnetic bias field $\bm{\mathrm{B}}_{\mathrm{T}} (t)$ that oscillates at kHz frequencies, which is fast compared to the mechanical response of the atoms but slow with respect to the rf field. By carefully controlling the relative influence of each field on the atom, TAAPs can be used to realise a broad palette of trap geometries where often simple, adiabatic routes to load and dynamically adjust the traps exist.

In this experiment the magnetostatic component is given by a quadrupole field of the form,
\begin{equation}
\bm{\mathrm{B}}_{\mathrm{q}} (\bm{\mathrm{r}}) = B'_{\mathrm{q}}(x \hat{\bm{e}}_{x}+y \hat{\bm{e}}_{y}-2z \hat{\bm{e}}_{z})
\label{bquad}
\end{equation}
where $B'_\mathrm{q}$ is the radial quadrupole gradient. Note that this field has its symmetry (z) axis aligned with gravity. A rf field that drives transitions between neighbouring Zeeman sub-states of a hyperfine manifold strongly modifies the potential experienced by atoms in the region of the quadrupole trap where the applied frequency $\omega_{\mathrm{rf}}$ equals the atomic Larmor frequency $\omega_{0} (\bm{\mathrm{r}}) =  |g_F \mu_{\mathrm{B}}\bm{\mathrm{B}} (\bm{\mathrm{r}})/\hbar|$.
The resonance condition $\omega_{0} (\bm{\mathrm{r}}) = \omega_{\mathrm{rf}}$ is satisfied along an iso-B surface, which in a cylindrically symmetric quadrupole describes an oblate spheroid centred on the point where $\bm{\mathrm{B}}_{\mathrm{q}} (\bm{\mathrm{r}}) = 0 $.
In the dressed atom picture, the rf field acts to induce an avoided crossing between the dressed states (these are eigenstates of the global system of atom + rf photon + interaction) where atoms that begin in the appropriate $m_F$ state can be trapped. In the absence of factors such as gravity or interaction the minimum in this adiabatic potential lies on the elliptical surface defined by the resonance condition. The dressed atom potential can be written as,
\begin{equation}
U(\bm{\mathrm{r}}) = m_F \hbar \sqrt{\delta^2(\bm{\mathrm{r}}) + \Omega_{\mathrm{R}}^2(\bm{\mathrm{r}})}
\end{equation}
where $\delta (\bm{\mathrm{r}}) = \omega_{0}(\bm{\mathrm{r}}) - \omega_{\mathrm{rf}}$ is the angular frequency detuning. The Rabi frequency of the rf transitions, $\Omega_{\mathrm{R}}$, quantifies the degree to which the atoms interact with the oscillating magnetic field and so is commonly referred to as the coupling strength. This quantity depends on the component of the oscillating field that is perpendicular to the \textit{local} magnetic field direction, and so is sensitive to the amplitude and \textit{polarisation} of the rf field. A general expression of the rf field in this experiment is given by,
\begin{eqnarray}
\bm{\mathrm{B}}_{\mathrm{rf}}(\bm{r},t) = &B^{x}_{\mathrm{rf}} \cos{(\omega_{\mathrm{rf}}t)}\hat{\bm{e}}_{x}+ B^{y}_{\mathrm{rf}} \sin{(\omega_{\mathrm{rf}}t)}\hat{\bm{e}}_{y}+& \nonumber \\
&B^{z}_{\mathrm{rf}} \cos{(\omega^{z}_{\mathrm{rf}}t+\alpha)}\hat{\bm{e}}_{z}&.
\label{brf}
\end{eqnarray}
The spatial dependence of $\Omega_{\mathrm{R}}$ in a quadrupole field for three specific polarisations of the applied rf field is shown in fig. \ref{polarisations} (Cases for which $B^{z}_{\mathrm{rf}} \neq 0$ are treated in sec.~\ref{vortexnuc} and illustrated in fig.~\ref{fig:rfrotation}).

The coupling strength is written as,
\begin{equation}
\Omega_{\mathrm{R}}(\bm{\mathrm{r}}) = \left| \frac{g_F \mu_{\mathrm{B}}}{2 \hbar} \frac{\bm{\mathrm{B}} (\bm{\mathrm{r}})}{|\bm{\mathrm{B}} (\bm{\mathrm{r}})|} \times \bm{\mathrm{B}}_{\mathrm{rf}} \right|.
\end{equation}
In the regions around the ellipsoidal minimum where the atoms couple strongly to the rf field (i.e. where there is a large component of the oscillating field $\bm{\mathrm{B}}_{\mathrm{rf}}$ perpendicular to the local static magnetic field) the potential is deformed by the mutual repulsion of the dressed states.

The technique of rf-dressing solves the problem of Majorana spin flips at the centre of a quadrupole trap by confining atoms on an elliptical surface which is always a minimum distance $r_{\mathrm{z}} = \hbar \omega_{\mathrm{rf}}/(2 m_{F} g_{F} \mu_{\mathrm{B}} B_{\mathrm{q}})$ (i.e. the semi-minor axis of the ellipsoid) away from where $\bm{\mathrm{B}}_{\mathrm{q}}(\bm{r}) = 0$. Despite this, atoms can still be rapidly lost from the TAAP if they make transitions to other (untrapped) dressed states as they traverse the avoided crossing. The rate of these diabatic Landau-Zener transitions is strongly dependent on the magnitude of the coupling strength i.e. the LZ loss channel is suppressed in the regions where $\Omega_{\mathrm{R}}(\bm{r})$ tends towards its maximum value and the repulsion between the dressed states is at its largest.  

The spatial inhomogeneity of the 3D quadrupole field means that there will always be at least one axis along which the local field $\bm{\mathrm{B}}_{\mathrm{q}}(\bm{r})$ and rf-dressing field are co-linear, resulting in zero coupling between the atoms and the rf field. The intersection of this axis with the resonant ellipsoid marks the position of `holes' in the adiabatic potential where the high LZ transition rate causes rapid atom loss from the trap. As is shown in Section \ref{natevap}, careful control of the potential can allow the LZ loss channel to be used to evaporatively cool an atomic sample across the BEC phase transition. A detailed understanding of how the TAAP potential can be altered by changing the polarisation of the rf-dressing field underlies the rotation scheme presented in Section \ref{vortexnuc}.

The magnitude and direction of the time-averaging field $\bm{\mathrm{B}}_{\mathrm{T}} (t)$ has a large influence on the resultant TAAP trap geometry. The role of the time-averaging field is to determine, in combination with the variation of the coupling strength and the force of gravity, the regions of the ellipsoid that the atoms access. In previous work a ring shaped potential was formed using a time-averaging field oscillating along the vertical ($z$) axis \cite{ben}. Starting from the same rf-dressed quadrupole, the vertically offset double-well potential used throughout the work described in this article is realised using a rotating bias field of the form,
\begin{equation}
\bm{\mathrm{B}}_{\mathrm{T}} (t) = B_{\mathrm{T}} \left[ \cos{(\omega_{\mathrm{T}}t)} \hat{\bm{e}}_{x} +\sin{(\omega_{\mathrm{T}}t)} \hat{\bm{e}}_{y} \right].  
\end{equation}
The time-averaging field shifts the centre of the adiabatic potential by an amount $r_{0} = B_{\mathrm{T}}/B'_q$ (in this case $r_{0}$ equals the radius of the circle of death in a conventional TOP trap \cite{petrich1995}). Therefore the adiabatic potential becomes $U_{\mathrm{AP}}(x + r_{0} \cos{\omega_{T}t}, y + r_{0}\sin{\omega_{T}t}, z)$. The full TAAP is found by evaluating the time-averaging integral,
\begin{equation}
U_{\mathrm{TAAP}} =\frac{\omega_{\mathrm{T}}}{2 \pi} \int^{\frac{2 \pi}{\omega_{\mathrm{T}}}}_{0}{U_{\mathrm{AP}}(\bm{r},t)} \mathrm{d}t.
\label{utaap}
\end{equation}

\section{Experimental Apparatus}
\label{exp}

All experimental data presented in this paper were gathered using our TOP trap apparatus which is configured to trap $^{87}\mathrm{Rb}$ in the $|F = 1, m_F = -1 \rangle$ hyperfine substate. Much of the versatility of this experiment derives from the ability to simultaneously deliver time-averaging and rf-dressing fields along all three mutually perpendicular axes, with independent control over the amplitude, frequency and phase of each field. This is facilitated by the pairing of a bespoke coil array with a custom designed direct digital synthesis (DDS) frequency source that can be updated at sub-millisecond intervals.

The vertically offset double-well TAAP trap is loaded from a standard radial TOP trap ($B'_{\mathrm{q}} = \unit[84]{G/cm}$, $B_{\mathrm{T}} = \unit[3.2]{G})$ in which \unit[$5 \times 10^6$]{atoms} are cooled to a temperature of \unit[1]{$\mu\mathrm{K}$}.
\begin{figure}[h]
\centering
\includegraphics[trim = 0mm 0mm 0mm 0mm, clip, scale=0.25]{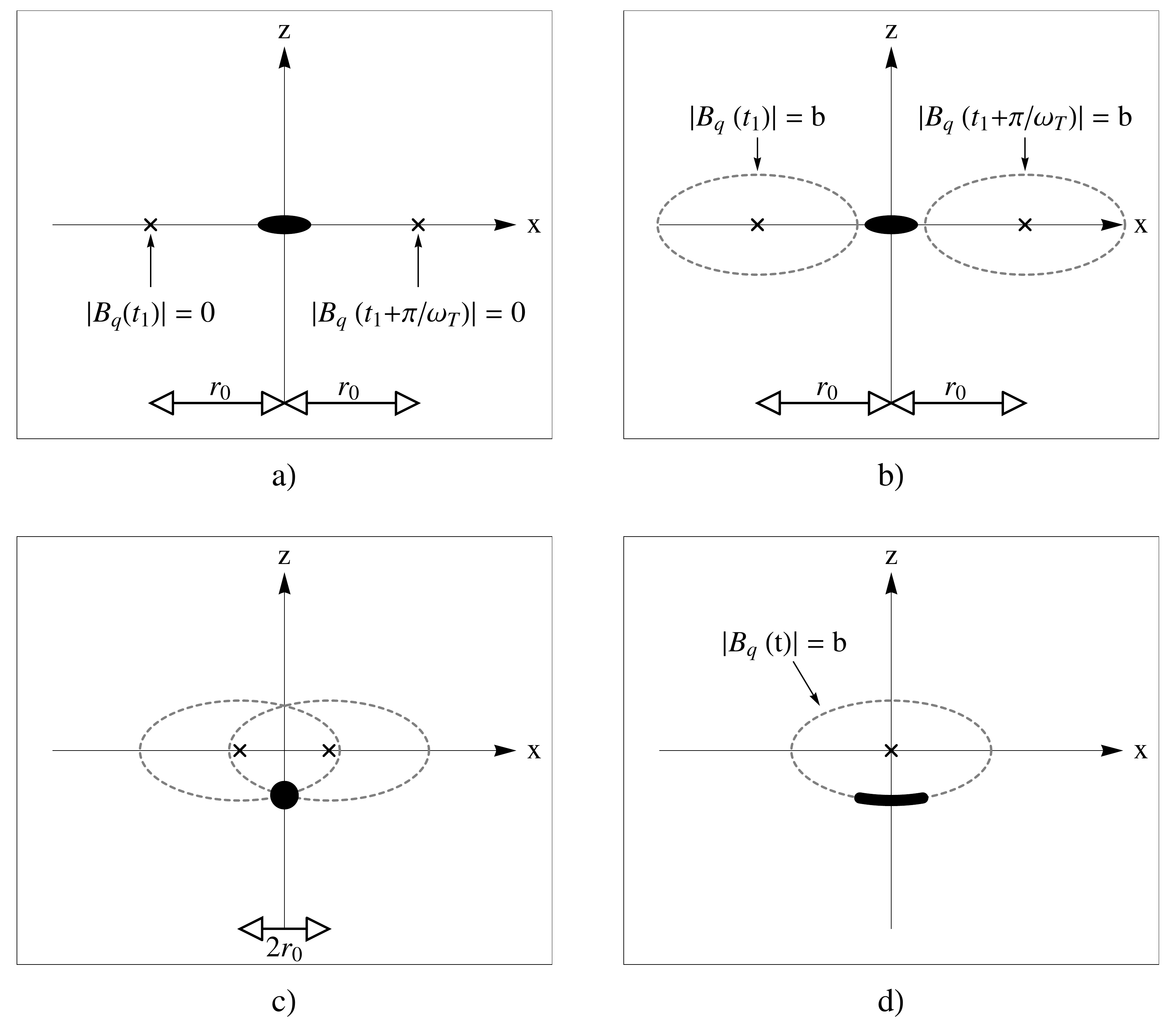}
\caption{Loading the vertically offset double-well TAAP trap. a) In a conventional TOP trap, under the action of a time-averaging field rotating in the $xy$ plane the zero of the quadrupole field $\mathrm{B_{q}(\bm{r})}$ is biased away from the position of the atoms (black) by a distance $r_0$ (shown at two times half an oscillation period apart). b) The application of a rf field forms a dressed state avoided crossing on an elliptical surface defined by the resonance condition (dashed line). Provided ellipsoid orbits exterior to the position of the atoms the TAAP closely resembles that of the TOP trap. c) The confinement experienced by the atoms is significantly modified when the amplitude of the bias field $B_{\mathrm{T}}$ is reduced such that the atoms locate on the avoided crossing. In this configuration, two minima form where the ellipsoid intersects the rotation axes. The vertical offset between them means that gravity breaks the symmetry of the wells and the atoms are loaded into the lower well. d) Ramping the time-averaging field to zero loads the atoms into the rf-dressed shell potential.}
\label{tlding}
\end{figure}
\begin{figure*}
\centering
\includegraphics[trim = 2mm 2mm 2mm 2mm, clip, width=0.45\columnwidth]{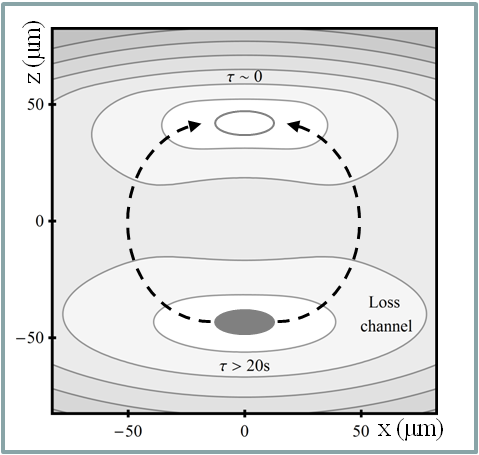}\hspace{0.0cm}
\includegraphics[width=0.8\columnwidth]{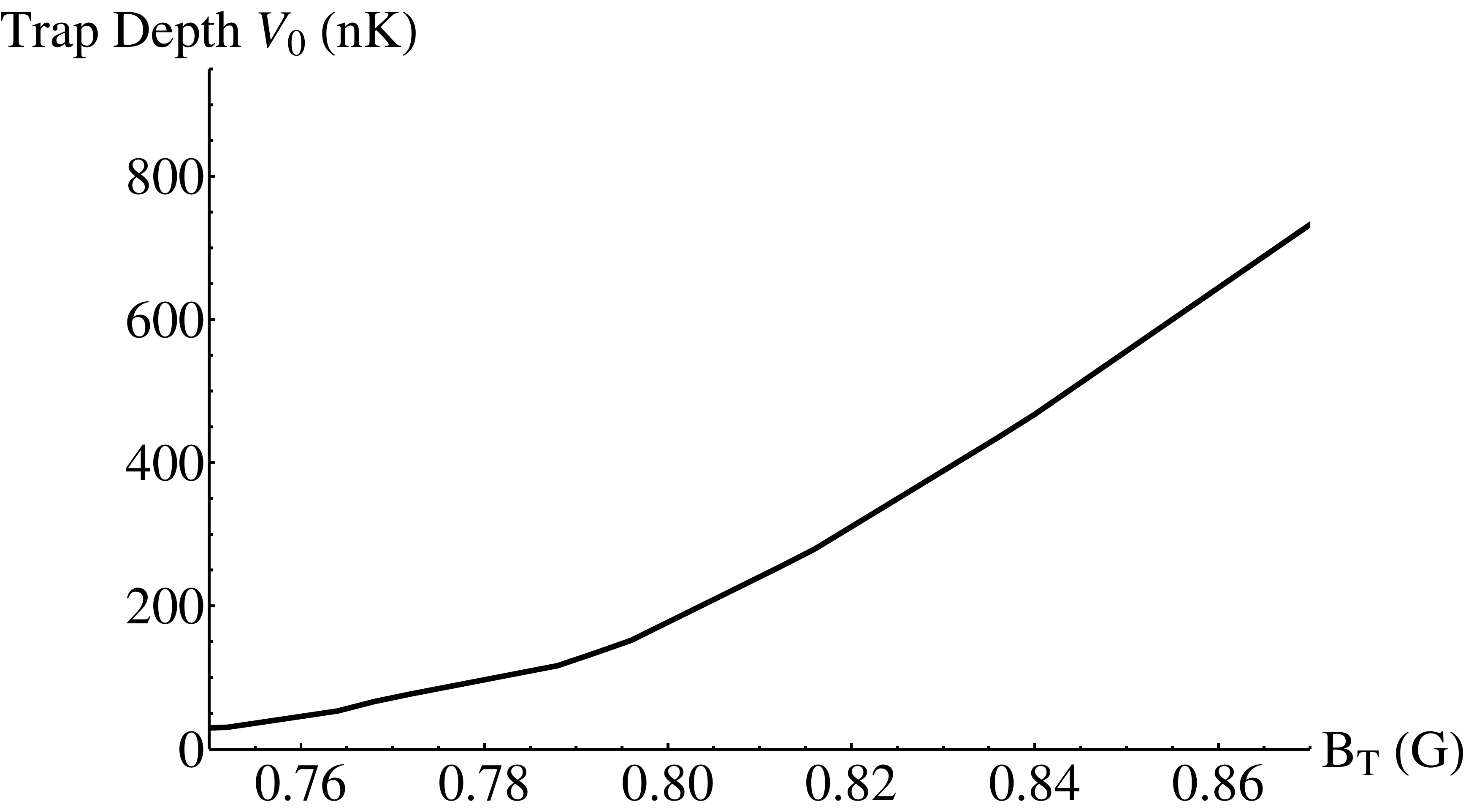}
\includegraphics[width=0.72\columnwidth]{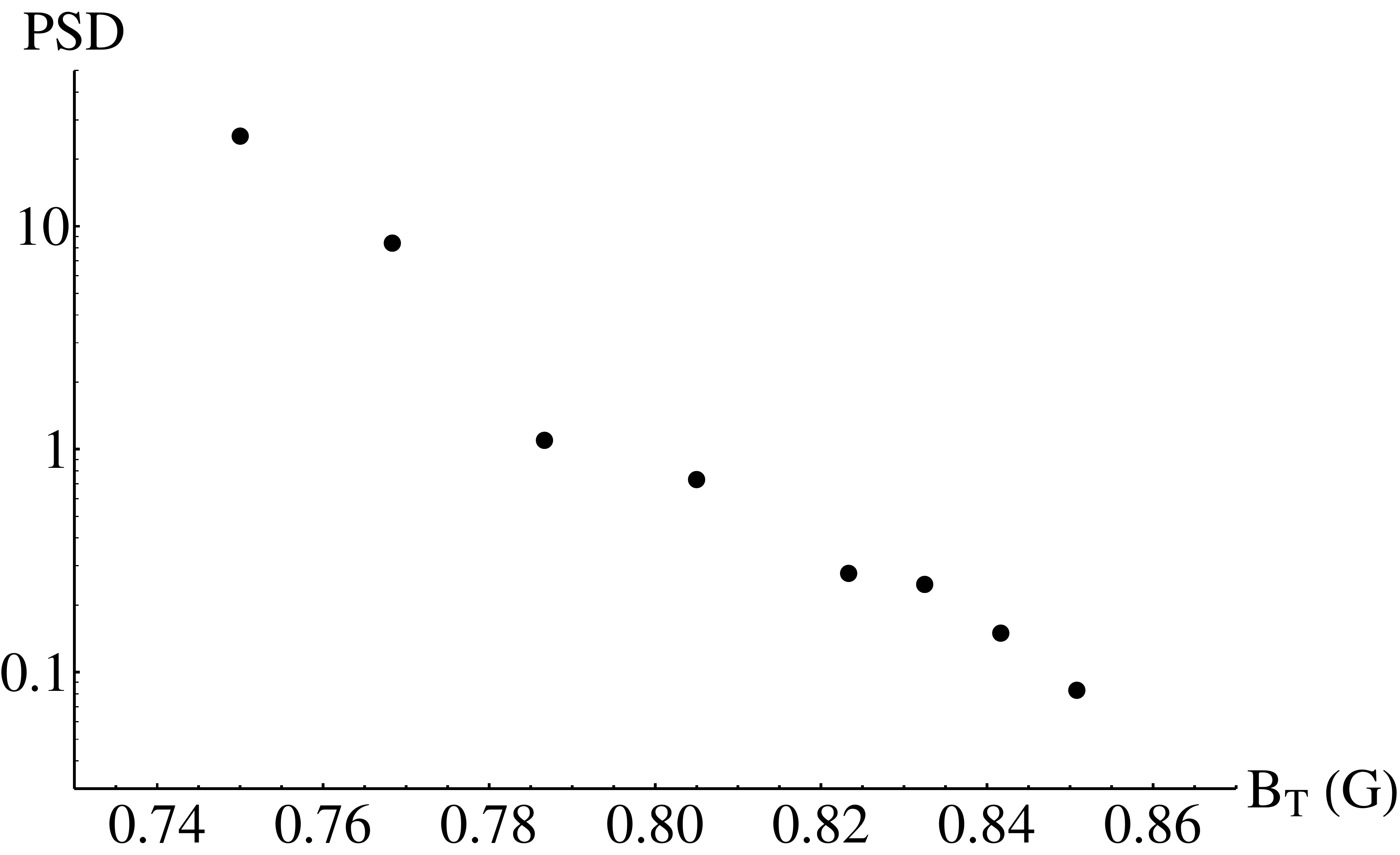}
\caption{Left: Contour plot of $U_{\mathrm{TAAP}}$ (see eqn.~\ref{utaap}), the vertically offset double-well TAAP formed with a circularly polarised ($\sigma^{+}$) rf-dressing field. Natural evaporation occurs when the most energetic atoms are able to flow along the indicated path from the lower to upper well site. The polarisation of the dressing field causes the lifetime in the upper well to be vanishingly short. Centre: Numerically generated plot showing the dependence of the trap depth $V_0$ in the lower well with $\mathrm{B}_{\mathrm{T}}$. Right: Measured increases in the phase space density of the trapped atomic gas as $\mathrm{B}_{\mathrm{T}}$ is ramped down.}
\label{trapdepth}
\end{figure*}
The loading sequence starts by switching on a circularly polarised rf-dressing field
\begin{equation}
\bm{\mathrm{B}}_{\mathrm{rf}}(\bm{\mathrm{r}},t) = B_{\mathrm{rf}}[\cos{(\omega_{\mathrm{rf}}t)}\hat{\bm{\mathrm{e}}}_{x}+\sin{(\omega_{\mathrm{rf}} t)}\hat{\bm{\mathrm{e}}}_{y}] 
\end{equation}
at fixed frequency ($\omega_{\mathrm{rf}} = \unit[2 \pi \times 1.4]{MHz}$) and amplitude ($B_{\mathrm{rf}} = \unit[0.7]{G}$). This field produces a radially symmetric coupling strength pattern that has a maximum at the resonant ellipsoid South pole and tends to zero at the North pole (see fig.~\ref{polarisations}a). The potential experienced by the atoms is not yet significantly modified by the dressing field as the amplitude of the TOP field ensures the resonant ellipsoid rotates exterior to the position of the atoms (see fig.~\ref{tlding}b). By ramping down $B_{\mathrm{T}}$ the resonant ellipsoid spirals inwards such that when $B_{\mathrm{T}} = \hbar \omega_{\mathrm{rf}}/(m_{F} g_{F} \mu_{\mathrm{B}}) =\unit[2]{G}$ the equatorial regions impinge on the atoms' position and the TAAP trap is loaded. The atoms are now trapped at the region where the elliptical surface intersects with the vertical rotation axis (fig.~\ref{tlding}c). To complete the loading sequence, the TOP field is typically ramped to $B_{\mathrm{T}} \leq \unit[1]{G}$ to ensure the atoms are trapped in regions of the ellipsoid where the coupling strength is sufficiently large to suppress LZ losses and give a sufficiently long trap lifetime.
\section{Natural evaporation}
\label{natevap}



Natural evaporation in the double-well TAAP trap exploits the hole in the rf-dressed potential (where $\Omega(\bm{r}) \rightarrow 0$, and therefore rapid Landau-Zener losses exist) to facilitate cooling of a trapped atomic sample. The potential is adjusted so that the energetic barrier separating the atoms' trapping position from the region where LZ losses occur rapidly is of the order of the thermal energy of the cloud. By slowly deforming the potential and lowering the trap depth, forced evaporative cooling is effected in a manner analogous to evaporative cooling in optical dipole traps \cite{chu1995}. 

To observe natural evaporation a cloud of $\unit[5 \times 10^5]{atoms}$ at a temperature of \unit[270]{nK} and a phase space density of 0.08 is prepared in the lower well of the vertically offset double-well TAAP trap, where $B'_{\mathrm{q}} = \unit[50]{G/cm}$ and $B_\mathrm{T}= \unit[1]{G}$. The rf-dressing field is circularly polarised such that $\Omega_{\mathrm{R}}(\bm{r}) \rightarrow 0$ at the North pole of the ellipsoid (see fig.~\ref{polarisations}~a). In this configuration, the spatial dependence of the coupling strength causes the two minima of the TAAP to pertain to very different lifetime characteristics. The upper well, which lies in the proximity of the hole in the rf-dressed potential, possesses a vanishing lifetime, while atoms can remain trapped at the lower well for over \unit[20]{s}. The loss channel for the process of natural evaporation is given by the path of least action between these two regions. We define the trap depth $V_0$ as the difference between the maximum value of the potential along this path and the value at the lower well.

Natural evaporation is brought about by controllably lowering the trap depth $V_0$, which in this experiment was performed in a two stage process. First the quadrupole gradient is adiabatically increased to $B'_q = \unit[176]{G/cm}$ over \unit[1]{s}. This decreases the size of the resonant ellipsoid, reducing the gravitational potential difference between the well sites and increases the trap confinement, increasing the elastic collision rate and thus improving conditions for cooling.

In the second stage the amplitude of the rotating bias field is ramped down across a narrow interval. The time-averaging field acts to limit the regions of the AP that are accessible to the atoms. As the amplitude $B_{\mathrm{T}}$ is reduced, its influence over the atoms weakens, allowing them to flow further from the lower well site. In the limiting case where $B_{\mathrm{T}} \rightarrow 0$, for this rf polarisation and quadrupole gradient no potential minimum exists at the lower well site.

The fine control over $B_{\mathrm{T}}$ allows the trap depth $V_0$ to be carefully adjusted such that those atoms with above average kinetic energy can escape to the upper well site (where they are lost) resulting in the overall cooling of the remaining sample. In this experiment, natural evaporative cooling across the phase transition is realised using a ramp in $B_{\mathrm{T}}$ from \unit[0.86]{G} to \unit[0.75]{G} over \unit[1.2]{s}, resulting in a BEC of $\unit[2.5 \times 10^5]{atoms}$ with no discernable thermal component. Further reductions to $B_{\mathrm{T}}$ below \unit[0.7]{G} result in complete atom loss from the trap. The efficiency of this process compares well with rf forced evaporation, where from similar starting values of temperature and phase space density a \unit[3]{s} evaporative sweep routinely produces a condensate of $\unit[3 \times 10^5]{atoms}$ in the TAAP trap.

The results presented in this section have a greater significance than simply an alternative to conventional evaporative cooling methods. They demonstrate the advances that are being made in the understanding of adiabatic potentials. The LZ loss channel was initially regarded as a significant obstacle for trapping cold atoms at energy level avoided crossings. This work has shown how this process can be harnessed in a beneficial way in future experiments in TAAPs.
\section{Vortex Nucleation in the TAAP trap} \label{vortexnuc}

The observation in this work of quantised vortices in the BEC is a potent demonstration of the power and versatility of the TAAP trap. The following section details how, via small changes to the rf-dressing field, an elliptical potential that rotates at a well defined frequency in the $xy$ plane is produced. The novel rotation scheme relies on the same mechanism for nucleating vortices that underlies some of the earliest experiments in this field \cite{hodby2001}. Under this mechanism, an elliptical deformation is introduced to a circularly symmetric trap and then rotated at the quadrupole surface mode frequency $\omega_q = \sqrt{2} \omega_r $. Following several hundred milliseconds of rotation, strong shape oscillations are observed in the cloud, which after a further few hundred milliseconds allow vortices to enter the cloud.

\begin{figure}[h]
\centering
\includegraphics[width=0.48\columnwidth]{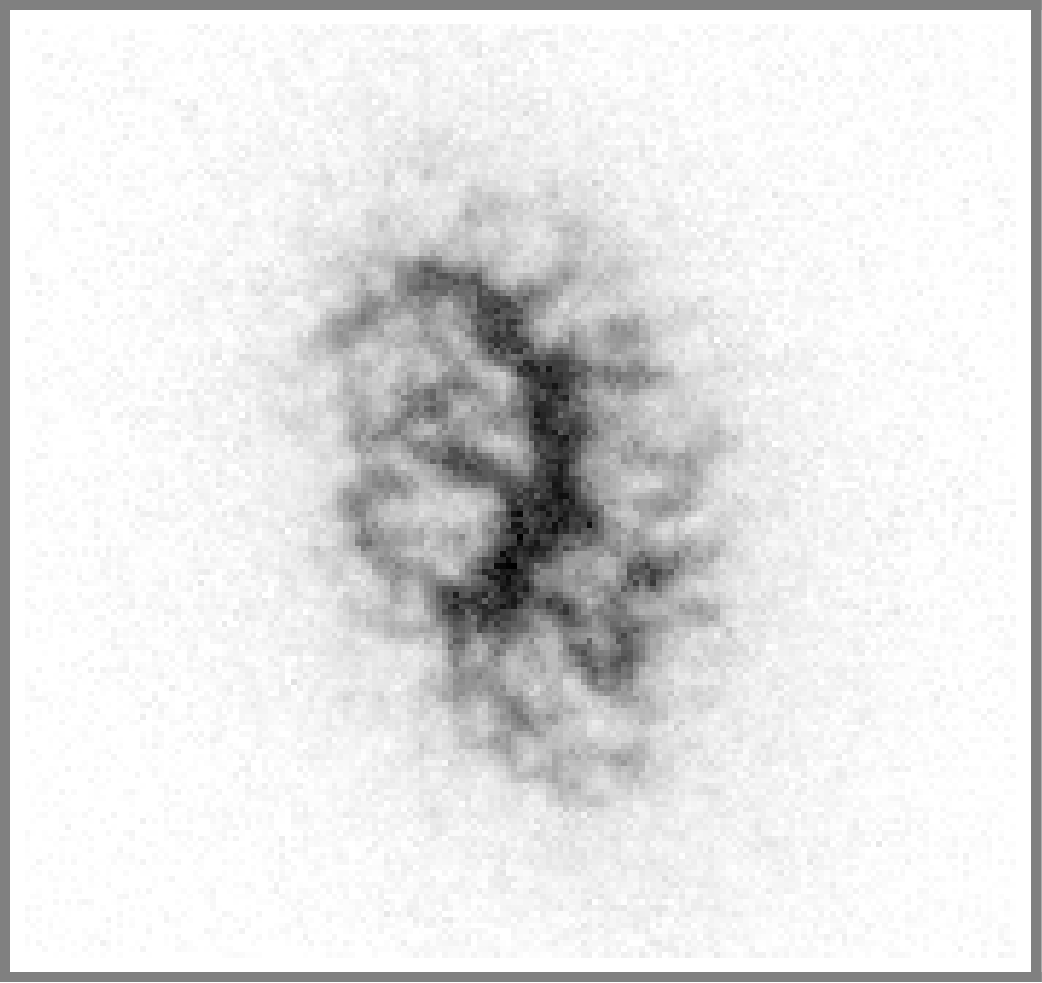}
\includegraphics[width=0.48\columnwidth]{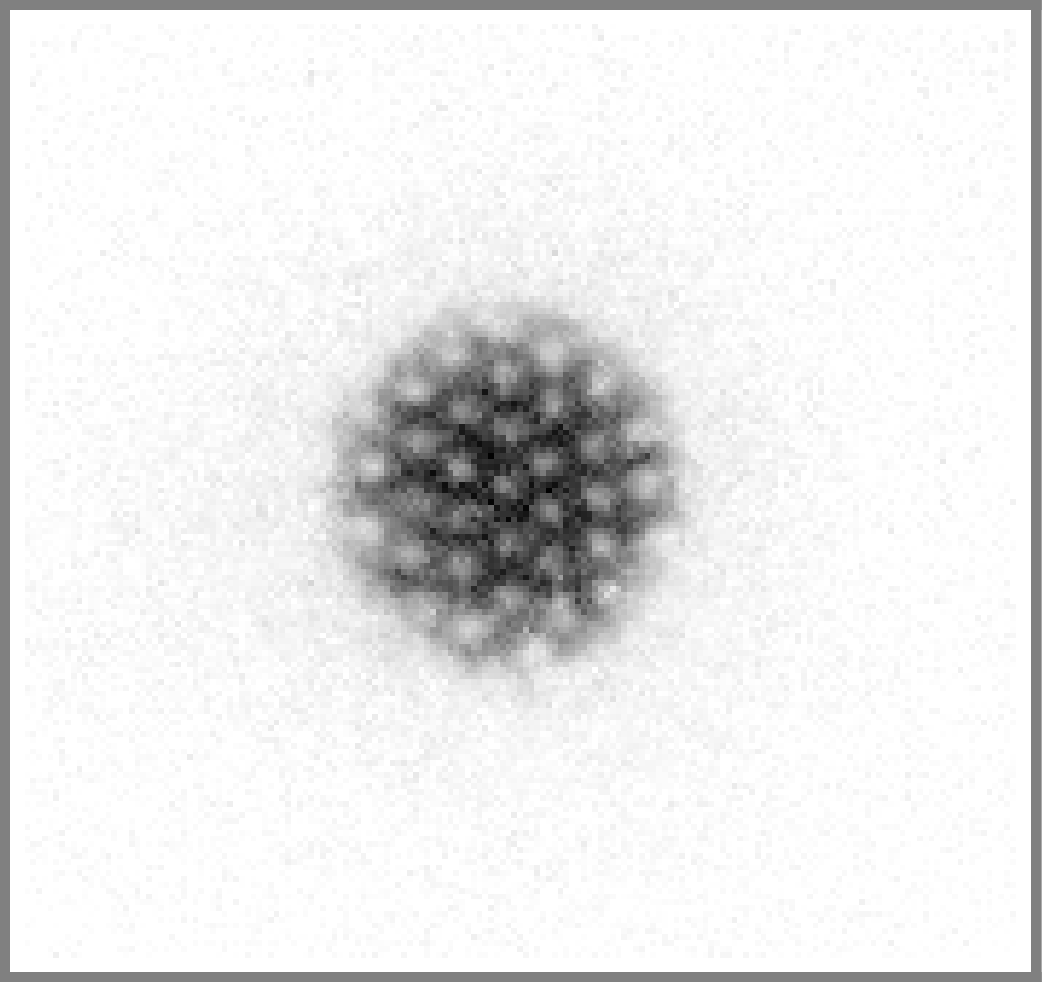}
\caption{Absorption images taken in the $xy$ plane after \unit[16]{ms} time of flight, showing vortex nucleation in the TAAP (extract sizes are \unit[$208 \times 208$]{$\mu \mathrm{m}^2$}). Left:~\unit[600]{ms} after introduction of the rotating elliptical deformation the cloud is highly distorted, allowing vortices to enter. Right:~\unit[2]{s} after returning to a circularly symmetric trap, a regular vortex array has formed.} \label{fig:vortexlattice}
\end{figure}

In \cite{hodby2001} the trap deformation and rotation was achieved by modulating the TOP field. In the present work the same manipulations occur by dynamically changing the polarisation of the rf-dressing field. By means of an overview, the following analysis will consider the effect on a bare quadrupole potential dressed with a $xy$ circularly polarised rf field, of introducing a third rf component applied along the vertical ($z$) axis. The impact on the potential of detuning the $z$ component from the $xy$ fields will be discussed before this system is considered together with the rotating bias field of the TOP trap.  

\begin{figure}[h]
\centering
\includegraphics[width=0.46\columnwidth]{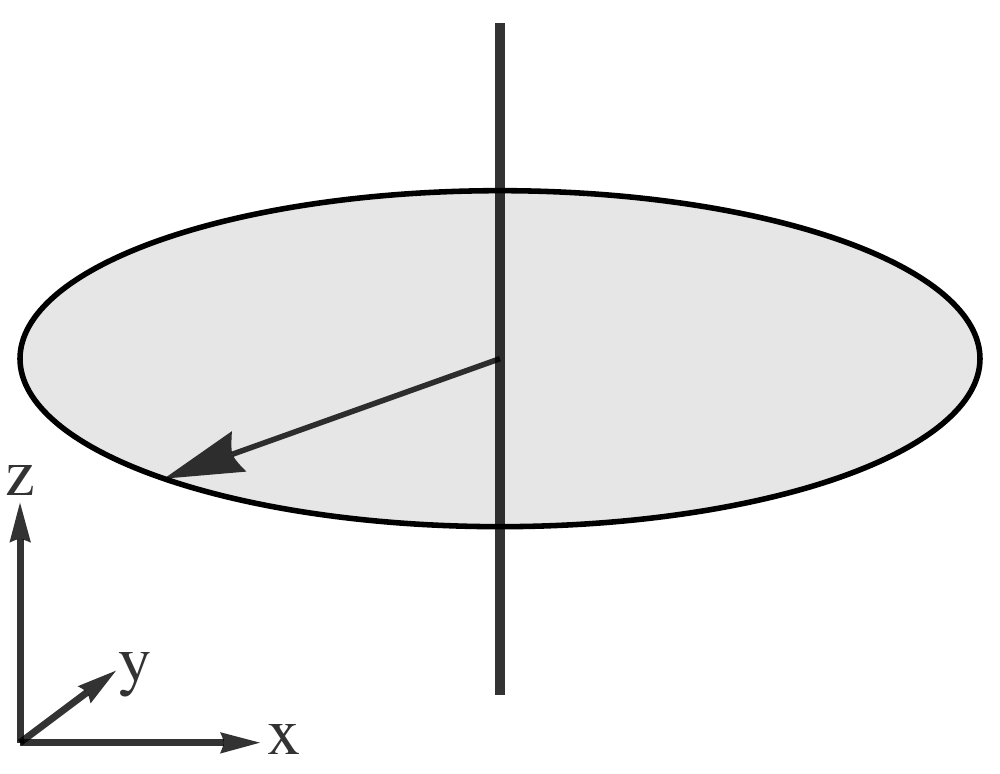}
\includegraphics[trim = 00mm 10mm 00mm 10mm, clip, width=0.52\columnwidth]{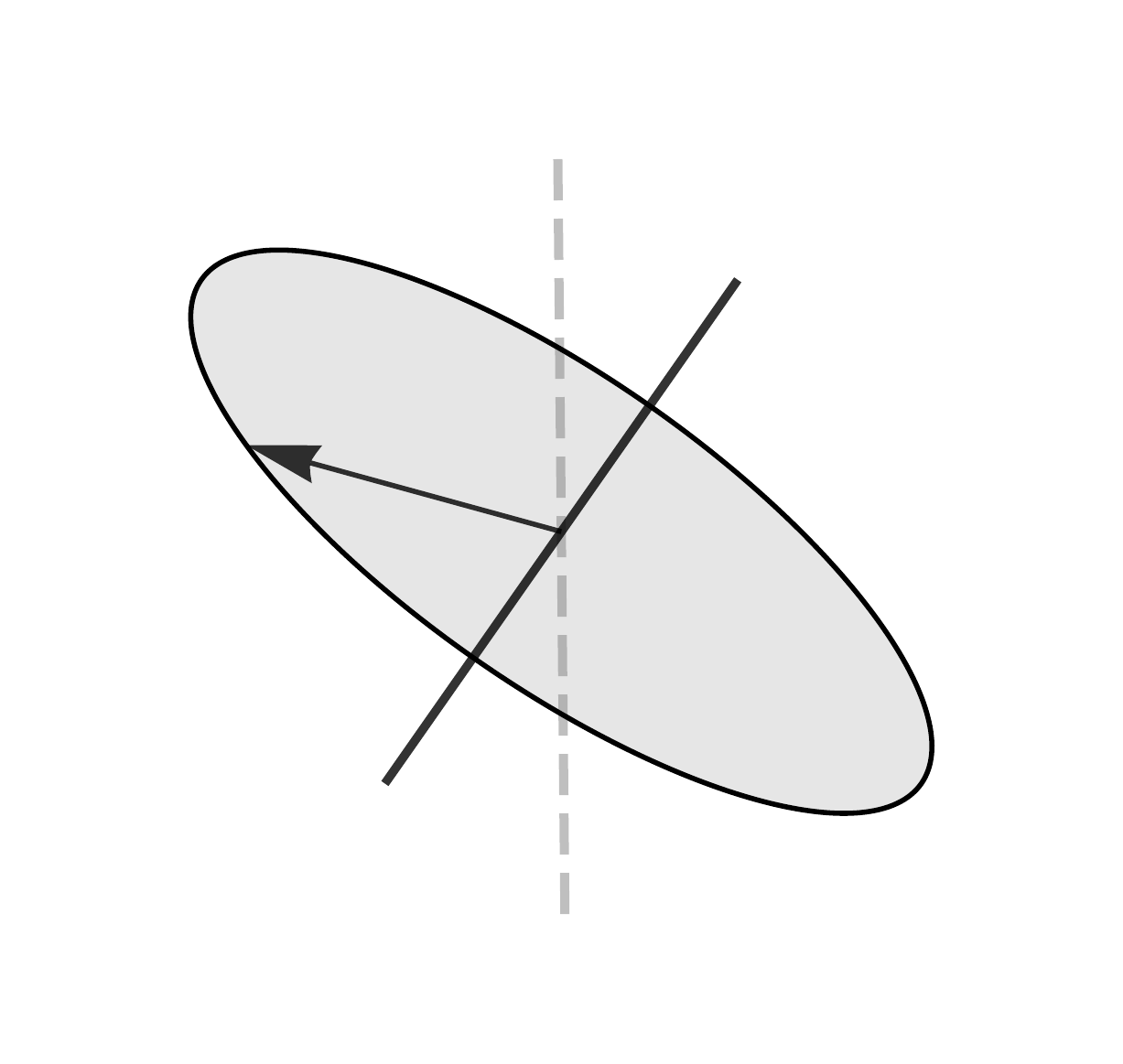}
\includegraphics[width=0.96\columnwidth]{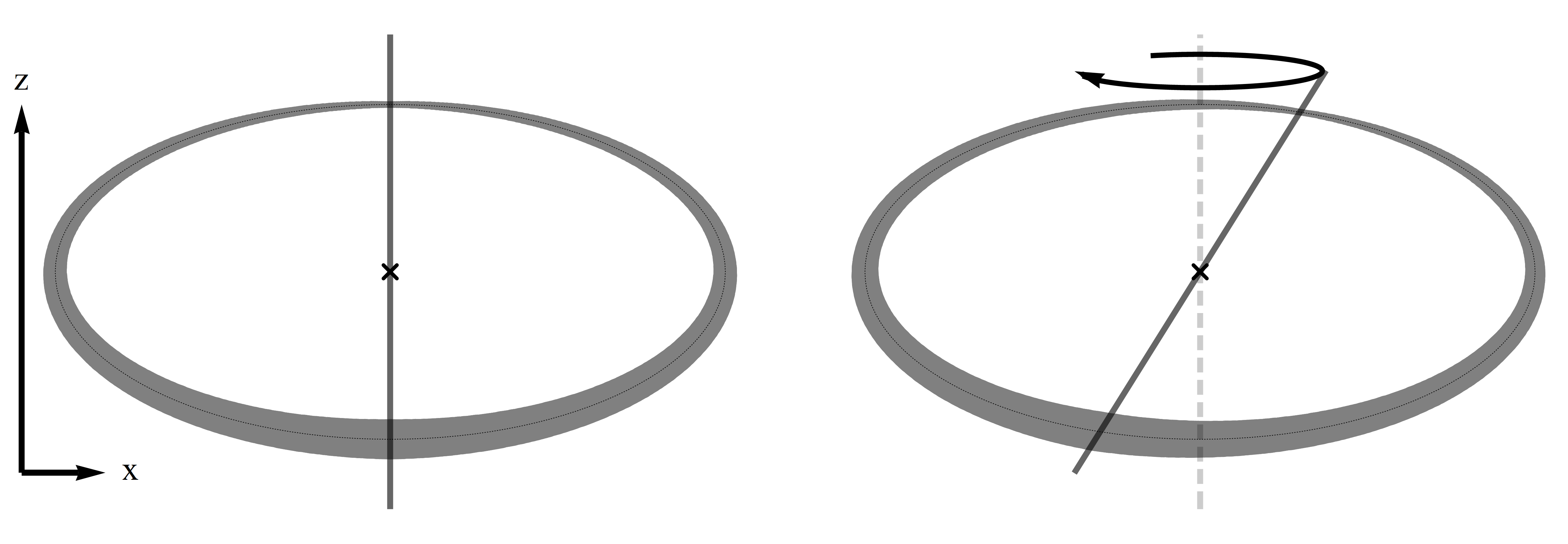}
\caption{Tilting and rotating the rf rotation axis. Top left: The gray disk depicts the area swept out by the rf field vector. Top right: When $\mathrm{B}_{\mathrm{rf}}^z \neq 0$ in eqn. \ref{brf} the rf rotation axis is tilted away from the z-axis. Bottom Left: Coupling strength distribution around the resonant ellipsoid for pure $xy$ circularly polarised ($\sigma^{+}$) rf field. Bottom right: The tilt in the rf rotation axis moves the region of maximum coupling away from the South pole of the resonant ellipsoid. If the $z$ component is detuned from the $x$ and $y$ components of the rf field, then the position of maximum coupling strength will rotate about the lower hemisphere of the ellipsoid.}
\label{fig:rfrotation}
\end{figure}

The spatial dependence of the coupling strength $\Omega_{\mathrm{R}}(\bm{r})$ in a quadrupole trap dressed by a rf field that is circularly polarised in the $xy$ plane is shown in fig.~\ref{polarisations}a. In this case the rf rotation axis (defined as the vector perpendicular to the plane of the rf field vector) is collinear with the $z$ axis (see fig.~\ref{fig:rfrotation}). The introduction of $\mathrm{B}_{\mathrm{rf}}^z$ at $\omega_{\mathrm{rf}}$ tilts the plane of rotation of the rf field vector and thus shifts the position where the coupling strength is a maximum away from the polar regions of the resonant ellipsoid. The direction of the tilt depends on the phase of the $z$-component of the rf-dressing field denoted by $\alpha$ in eqn. \ref{brf}. This concept was applied in \cite{ben} to improve the balancing of a horizontally orientated ring trap. Further by introducing a small frequency difference between the $z$ and $xy$ components of the rf-dressing field such that $\delta \omega_{\mathrm{rf}} =  \omega_{\mathrm{rf}} - \omega_{\mathrm{rf}}^z \neq 0$ (and $\delta \omega_{\mathrm{rf}} \ll \omega_{\mathrm{rf}}$), causes the rotation axis of $\mathrm{B}_{\mathrm{rf}}(\bm{r},t)$ to precess about the $z$ axis at $\delta \omega_{\mathrm{rf}}$ (see bottom right of fig.~\ref{fig:rfrotation}).

In the rf-dressed shell potential ($B_\mathrm{T}=0$) the result is that the maximum of the coupling strength orbits around the South pole at a frequency equal to $\delta \omega_{\mathrm{rf}}$, with an orbital radius that is determined by the ratio $\mathrm{B^z_{\mathrm{rf}}}/\mathrm{B_{\mathrm{rf}}}$. We chose $\delta \omega_{\mathrm{rf}}$ to be large with respect to the trapping frequencies i.e. $\delta \omega_{\mathrm{rf}} \gg \omega_{x,z}$ so that the atoms experience the time-average of the coupling strength \cite{perrin2010}. This effect can be used to increase the strength of the confinement offered by the rf-dressed shell trap, and is in combination with the rotating bias field of the TOP trap, the basis for the rotation scheme.

\begin{figure}[h]
\centering
\includegraphics[width=0.99\columnwidth]{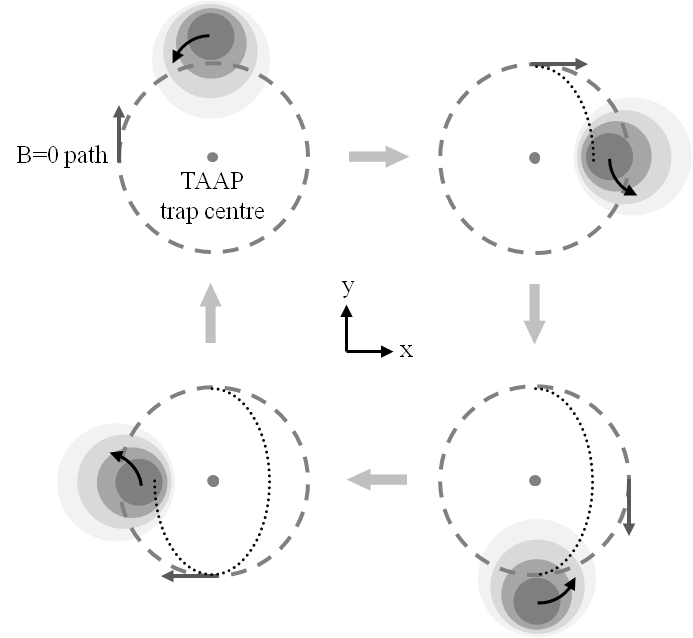}
\caption{Counter-rotating the rf rotation axis and the rotating bias field. Each subfigure shows the situation in the $xy$ plane of the TAAP trap at time intervals of $2\pi / 4 \omega_{\mathrm{T}}$. The shaded regions represent the variation of the coupling strength around its maximum position. The dashed line shows the locus of the quadrupole field zero under the action of the rotating bias field. The dotted line indicates the resulting elliptical trap geometry.} \label{fig:fullrotation} 
\end{figure}

We observe vortex nucleation in the TAAP trap, and so it is now necessary to combine the rf-dressed shell potential where the rf rotation axis precesses about the $z$ axis with rotating bias field from the TOP trap. The resulting potential contains two rotating components (the orientation of the rf rotation axis and the rotating time-averaging field) that can for a particular set of parameters give rise to a slowly rotating, elliptical trap. In order to visualise this, first consider the scenario where both components rotate at the same frequency ($\delta \omega_{\mathrm{rf}} = \omega_{\mathrm{T}}$) and with the same handedness. In this co-rotating case, the TOP field vector and the tilt direction of the rf rotation axis are spatially locked (i.e. they rotate with a constant phase difference). The resultant trapping potential is circularly symmetric (in the $xy$ plane).  
In the more pertinent counter-rotating case the symmetry in the $xy$ plane is broken and the trap adopts an elliptical geometry as shown in fig. \ref{fig:fullrotation}. Critically the relative phase between the two rotating components sets the orientation of the ellipse in the $xy$ plane. Consequently a changing relative phase causes the axes of the ellipse to rotate. Such a situation is straightforward to realise by setting $\delta \omega_{\mathrm{rf}} = \omega_{\mathrm{T}} \pm 2 \omega_{\mathrm{rot}}$ (and $\omega_{\mathrm{rot}} \ll \omega_\mathrm{T}$), where $\omega_{\mathrm{rot}}$ is the rotation frequency of the ellipse \footnote{Mathematically this can be seen from the path $\bm{r}(t)$ that is traced out by the addition of the two rotating vectors: 
\[
\bm{r}(t) = \bm{r_1}(t) + \bm{r_2}(t) = r_1 \left( \begin{array}{c} \cos (\omega_1 t) \\ \sin (\omega_1 t) \end{array}
\right)
+
r_2
\left(
\begin{array}{c}
\cos (\omega_2 t) \\
\sin (\omega_2 t)
\end{array}
\right)
\label{eq:rotellipse}
\]

With $\omega_2 = \omega_0 - \frac{\delta\omega}{2}$ and $\omega_1 = -\omega_0 + \frac{\delta\omega}{2}$ it can be shown that 
\[
\bm{r}(t) = R \left[ \frac{\delta\omega t}{2} \right] \cdot \left( \begin{array}{c} (r_1+ r_2 ) \cdot \cos (-\omega_0 t) \\
(r_1- r_2 ) \cdot \sin (-\omega_0 t)
\end{array}
\right)
\label{eq:rotellipse2}
\]
where $R$ is the rotation matrix. This describes the slow rotation of an elliptical path.
}.

\begin{figure}[h]
\centering
\includegraphics[width=0.99\columnwidth]{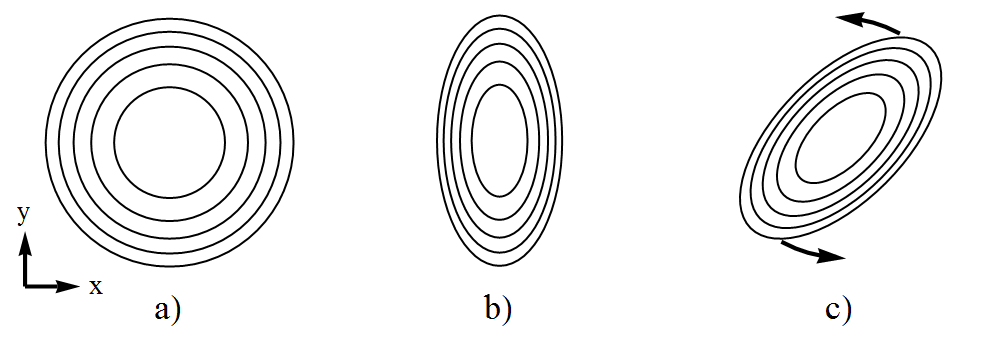}
\caption{Contour plots of the lower well of the TAAP in the $xy$ plane. a) The symmetry for a pure $xy$ circularly polarised rf-dressing field, is broken when b) a $z$-component of the dressing field is introduced at a counter-rotating frequency $\omega_{\mathrm{rf}} + \omega_{\mathrm{T}}$. If the frequency of this field is detuned slightly, the result is c) a rotating elliptical potential.} 
\label{clouds} 
\end{figure}

In this experiment $\omega_{\mathrm{T}}= 2\pi \times \unit[7]{kHz}$ and $\omega_{\mathrm{rf}} = 2\pi \times \unit[1.400000]{MHz}$, so that $\omega_{\mathrm{rf}}^z = 2\pi \times \unit[1.407000]{MHz}$ creates the described elliptical deformation of the TAAP trap. Detuning the value of $\omega_{\mathrm{rf}}^z$ to $\omega_{\mathrm{rf}}^z = 2\pi \times \unit[1.407028]{MHz}$ causes the elliptical trap minimum to rotate in the $xy$ plane at a frequency $\omega_{\mathrm{rot}} = 2\pi \times \unit[14]{Hz}$. This is close to the quadrupole frequency ($\omega_x = 2\pi \times \unit[20]{Hz}$) of our TAAP trap, and leads to the desired quadrupole mode excitation. Figure \ref{fig:vortexlattice} displays the results for a BEC of such a procedure.

Lattices of between 50 - 60 vortices are routinely realised using this technique and it is anticipated that higher numbers could be attained by refining the technique further. In general vortex nucleation in the TAAP behaves in a similar way to that which was previously reported in the TOP trap \cite{hodby2001}. The number of vortices depends upon the frequency difference between the rotating ellipsoid and the quadrupole mode, and also on degree of ellipticity of the trap. In all cases, a regular array of vortices only forms after the cloud is returned to a circularly symmetric trap and held in this potential for a settling period of $\sim \unit[1.5]{s}$. To improve the contrast and clarity of the vortex lattices, a rf evaporation field is applied at a constant frequency throughout the vortex nucleation sequence. 

\section{Conclusion and outlook}
\label{conc}

In conclusion evidence for the successful implementation of novel mechanisms to cool and rotate a cloud of ultracold atoms in a time-averaged adiabatic potential is presented. Natural evaporative cooling across the BEC phase transition using only the LZ loss channel and the nucleation of vortices via trap deformation and rotation were observed for atoms trapped in the lower well of the vertically offset double-well TAAP trap. Natural evaporation and vortex nucleation are significant milestones and further illustrate the versatility of the TAAP. The manipulations required to perform each process are fully realised by the adjustment of one or more of the three magnetic and rf fields that give rise to the TAAP. No other external potential is required. 

The TAAP continues to demonstrate its suitability for controlling and manipulating ultracold quantum gases. A new cooling technique and the ability to impart angular momentum to a BEC augment our previous work using the TAAP to realise non-trivial trapping geometries. A long term goal of the research community has been to observe and analyse the behaviour of ultracold atoms in the strongly correlated fraction quantum Hall (FQH) regime \cite{cooper2008, fetter2009}. The TAAP possesses several features that are highly desirable with this objective in mind. Versatile potentials allow the dynamics of the trapped atomic system, including as we have shown its angular momentum, to be adjusted with a high degree of accuracy. The ability to smoothly switch between ring-shaped and harmonic trap geometries has been proposed as a technique that mitigates the stringent experimental requirements for realising FQH states in ultracold gases ~\cite{dalibard2011}. Finally the very low energy scales of these system require low heating rates and the ability of actively cool the sample, both of which are provided by the TAAP. 

\section{Acknowledgments}
This work has been supported by the Engineering and Physical Science Research Council under the Grant No. EP/E010873/1.

\end{document}